\documentclass[a4paper,fleqn]{cas-sc}

\usepackage[authoryear,sort]{natbib}
\usepackage{chemformula}
\usepackage{amsmath}

\def\tsc#1{\csdef{#1}{\textsc{\lowercase{#1}}\xspace}}
\tsc{WGM}
\tsc{QE}
\tsc{EP}
\tsc{PMS}
\tsc{BEC}
\tsc{DE}

\begin{document}
\let\WriteBookmarks\relax
\def\floatpagepagefraction{1}
\def\textpagefraction{.001}

\shorttitle{Synthesis of urea on icy mantles}

\shortauthors{J. Perrero et~al.}

\title [mode = title]{Synthesis of urea on the surface of interstellar water ice clusters. A quantum chemical study.}

\author[1,2]{J. Perrero}[orcid=0000-0003-2161-9120]
\cormark[1]

\ead{jessica.perrero@uab.cat}

\affiliation[1]{organization={Departament de Química, Universitat
Autònoma de Barcelona}, 
    city={Bellaterra},
    postcode={08193}, 
    country={Catalonia, Spain}}


\author[1]{A. Rimola}[orcid=0000-0002-9637-4554]
\cormark[1]

\ead{albert.rimola@uab.cat}

\affiliation[2]{organization={Dipartimento di Chimica and Nanostructured Interfaces and
Surfaces (NIS) Centre, Università degli Studi di Torino},
    city={Torino},
    postcode={10125}, 
    country={Italy}}

\cortext[cor1]{Corresponding authors}

\begin{abstract}    
Urea is a prebiotic molecule that has been detected in few sources of the interstellar medium (ISM) and in Murchison meteorite. Being stable against ultraviolet radiation and high-energy electron bombardment, urea is expected to be present in interstellar ices. Theoretical and experimental studies suggest that isocyanic acid (HNCO) and formamide (NH$_2$CHO) are possible precursors of urea. However, uncertainties still exist regarding its formation routes. Previous computational works characterised urea formation in the gas phase or in presence of few water molecules by reaction of formamide with nitrogen-bearing species. In this work, we investigated the reaction of HNCO + \ch{NH3} on an 18 water molecules ice cluster model mimicking interstellar ice mantles by means of quantum chemical computations. We characterised different mechanisms involving both closed-shell and open-shell species at B3LYP-D3(BJ)/ma-def2-TZVP level of theory, in which the radical-radical H$_2$NCO + NH$_2$ coupling has been found to be the most favourable one due to being almost barrierless.
In this path, the presence of the icy surfaces is crucial for acting as reactant concentrators/suppliers, as well as third bodies able to dissipate the energy liberated during the urea formation.

\end{abstract}




\maketitle

\section{Introduction}

Nowadays, about 300 gas-phase molecules have been detected in the interstellar medium (ISM), out of which a few are amides, e.g., formamide, acetamide, N-methylformamide or urea \citep{belloche2017,database, ligterink2022, zeng2023}. The latter has the particularity to possess two \ch{C-N} bonds and is one of the key ingredients required in the synthesis of purines and pirimidines \citep{robertson1995,becker2019,menor2020}. These nucleobases are the building blocks of the nucleotides that constitute both the RNA and the DNA \citep{kolb2014astrobiology}. 
These characteristics make urea one of the key species in the field of astrochemistry and origin of life studies. 

Urea, \ch{NH2CONH2}, was first detected in the Murchinson meteorite \citep{hayatsu1975}, and more recently it was observed towards Sagittarius B2(N1) by \cite{belloche2019}, towards G+0.693-0.027 giant molecular cloud by both \cite{jimenez-serra2020} and \cite{zeng2023}, and tentatively detected by \cite{raunier2004detection} in NGC 7538 IRS9 and by \cite{Remijan2014} in Sagittarius B2(N-LMH). However, observations dedicated to determine the inventory of interstellar amides towards the calss 0 protostar SMM1-a in Serpens cloud \citep{ligterink2022} did not report the presence of urea.

Laboratory experiments carried out by \cite{herrero2022} on the irradiation of \ch{NH2CONH2} and \ch{NH2CONH2}:H$_2$O ices show that urea is stable against ultraviolet irradiation and high-energy electron bombardment in cold molecular clouds and hot cores, thereby suggesting that grain mantles could be a reservoir of urea in the ISM.

Despite the numerous studies on its synthetic pathways, there are still uncertainties regarding urea formation in the ISM. Previous experimental and theoretical investigations have proposed two precursors of urea: isocyanic acid, HNCO, and formamide, NH$_2$CHO. 

HNCO can be considered a molecule of prebiotic interest and it was first detected in 1972 in the ISM \citep{snyder1972}, in cometary comae \citep{winnewisser1999}, and in external galaxies \citep{nguyen1991}. \cite{jimenez2014} obtained solid HNCO by UV irradiation of H$_2$O:NH$_3$:CO and H$_2$O:HCN ice mixtures. A successive experimental study of \cite{nourry2015} suggested that the reaction of N + CO on water icy mantles could be responsible for its production. Alternatively, in translucent and dense clouds, HNCO (together with NH$_3$) was proposed to be produced by co-deposition of N, H and CO \citep{fedoseev2015}.

HNCO was first proposed as a urea precursor by \cite{raunier2004detection}, in which vacuum ultraviolet irradiation of pure HNCO ice at 10 K produced ammonium cyanate (\ch{NH4+ OCN-}), formamide and urea. The formation of ammonium cyanate can possess an activation barrier or be spontaneous depending on the chemical environment of the reactants. While \cite{raunier2003nh4ocn} described the spontaneous deprotonation of HNCO embedded in NH$_3$/H$_2$O ice mixtures, \cite{mispelaer2012} observed that thermal processing of HNCO:NH$_3$ ices is necessary to overcome the small barrier of the acid-base reaction. Nevertheless, the deprotonation of isocyanic acid could explain why it has never been detected in icy mantles, while the \ch{OCN-} feature has been documented \citep{boogert2015,mcclure2023}.

Another hypothesis postulates formamide as a precursor of urea \citep{forstel2016}, in which the irradiation of NH$_3$:CO ices first forms formamide and then urea. \cite{meijer2019} and \cite{slate2020} characterised by means of computational chemistry calculations the mechanisms proposed in \cite{forstel2016}. The calculations showed that the concerted reaction between formamide and ammonia, as well as the two reactions obtained by converting one of the reactants into a radical, are not feasible under ISM conditions. Thus, an alternative set of reactions involving radicals and charged species was proposed, both in the gas phase and in the presence of very few (e.g., three) water molecules as minimal cluster models mimicking an ice mantle. These pathways show no or very low barriers.  
Another set of reactions involving formamide was characterised by \cite{brigiano2018}, who studied several pathways to form urea in the gas phase. The majority of them possess high activation barriers, hence being unfeasible in the ISM. In this work, again, charged species provided a solution to the lack of energy input of the ISM, with HCONH$_2$ + \ch{NH2OH2+} being the most favourable reaction. 

Given the presence of a thorough computational study on formamide reactions, in this work we investigated the reactivity of HNCO and derivatives, focussing in particular on the effect that ice mantles can have on their reactions. Icy grains can play multiple roles in astrochemical reactions: (i) they can act as a chemical catalyst, reducing energy barriers and accelerating reactions \citep{potapov2021}, (ii) they can serve as reactant concentrators and even suppliers, and (iii) they can exert the third body effect, dissipating the energy released by exothermic processes, hence stabilising the products formed \citep[e.g.,][]{pantaleone2020,pantaleone2021,ferrero2023}. The presence of the ice surface can be crucial for the positive outcome of a reaction which, otherwise, would be unfeasible in the gas phase. For example, in previous theoretical works by us, we investigated the reactivity of a radical (CN and CCH) with a component of the ice mantle (H$_2$O), resulting in the synthesis of formamide and ethanol with small or no barriers \citep{rimola2018,perrero2022}. 

In the present work, we modelled different synthetic routes for urea formation using HNCO as reactant (instead of NH$_2$CHO, the main reactant in previous theoretical works), by simulating different processes on the surface of an 18 H$_2$O molecules cluster, a robust model to mimic interstellar water ice surfaces for chemical reactions \citep{Zamirri2019,Rimola2021}. 

We first considered the reactivity between HNCO and NH$_3$ (closed-shell/closed-shell reaction), that is:

\begin{equation}\label{eq:neu-neu}
  \ch{HNCO + NH3 -> NH2CONH2}  
\end{equation}

However, this reaction has a competitive channel, the formation of the \ch{NH4+ OCN-} ion pair by proton transfer from HNCO to NH$_3$:

\begin{equation}\label{eq:salt}
     \ch{HNCO + NH3 -> NH4+ OCN-}   
\end{equation}

The formation of the ion pair cannot be overlooked given that it can be favoured by the presence of the water ice surfaces, because of the ability of water to stabilise charged species.

As a second possible channel for the formation of urea, we also simulated the reactivity of HNCO with NH$_2$ (closed-shell/radical reaction), giving NHCONH$_2$ (an urea radical precursor), whose successive hydrogenation yields the actual urea molecule:

\begin{equation}\label{eq:rad-neu}
     \ch{HNCO + NH2 -> NHCONH2}  
\end{equation}   
\begin{equation}\label{eq:hydrogenation}
     \ch{NHCONH2 + H -> NH2CONH2}   
\end{equation}

Finally, we considered a third route, the radical-radical reaction between H$_2$NCO and NH$_2$. However, this can have two outcomes, depending on the orientation of the reactants: direct formation of urea (for which we also computed the kinetics) or formation of HNCO and NH$_3$ by direct H-transfer from H$_2$NCO to NH$_2$. That is:

\begin{equation}\label{eq:rad-rad}
     \ch{H2NCO + NH2 -> NH2CONH2}
\end{equation}     
\begin{equation}\label{eq:rad-rad-reagent}
     \ch{H2NCO + NH2 -> NH3 + HNCO} 
\end{equation}

In this work, we present the results obtained for the simulation of these reactions by means of quantum chemical calculations.

\section{Methods}\label{sec:method}

All calculations were performed with the ORCA (v.5.0.3) programme suite \citep{neese2022}. 
We ran geometry optimisations and frequency calculations with the hybrid DFT functional B3LYP \citep{lyp:1988,becke:1988,becke:1993} including the Grimme's D3 empirical correction using the Becke-Johnson (BJ) damping scheme \citep{grimme:2010,grimme:2011} and combined with the ma-def2-TZVP basis set \citep{zheng2011}. 

We identified the reactants and the products of each reaction, and, if necessary, the transition states through to the nudged elastic band (NEB) method. Frequency calculations were carried out to confirm the nature of the stationary points, namely, reactants and products as minima and transition states as first-order saddle points of the potential energy surfaces. B3LYP-D3(BJ)/ma-def2-TZVP electronic energies were corrected with the vibrational zero-point energy (ZPE) term for each stationary point, thus obtaining enthalpy variations at 0 K according to the equation, e.g., for energy barriers, $\Delta$H$_{TS}$ = $\Delta$E$_{TS}$ + $\Delta$ZPE.

To assess the accuracy of the B3LYP-D3(BJ)/ma-def2-TZVP methodology, we performed single point energy calculations on the optimised structures at a higher level of theory, that is, the `gold standard' of quantum chemistry CCSD(T) \citep{ccsdt} with the inclusion of explicitly correlated terms in the wave function to account for electron correlation effects: the so-called CCSD(T)-F12 method \citep[][and references therein]{adler2007}. The calculations are performed with a combination of correlation consistent basis sets, in our case cc-pVTZ-F12 as the main basis set, with cc-pVTZ-F12-CABS, aug-cc-pVTZ, and aug-cc-pVQZ/C as auxiliary basis sets. The benchmark has been done for reaction \ref{eq:neu-neu} both in the gas phase (i.e., absence of water) and in the presence of one water molecule.

Additionally, to improve the accuracy of the results obtained for the reactions occurring on the cluster model at a feasible computational cost, we used the domain based local pair natural orbital coupled cluster theory with single-, double-, and perturbative triple-excitations, DLPNO-CCSD(T) \citep{riplinger2016}, to which the same F12 approximation is added \citep{pavosevic2017}. This methodology was applied to the reaction \ref{eq:neu-neu} when simulated on the water ice cluster model. 

Reactions \ref{eq:rad-neu}, \ref{eq:hydrogenation}, \ref{eq:rad-rad}, and \ref{eq:rad-rad-reagent} involve structures that are open-shell in nature. Thus, the calculations were performed within the unrestricted formalism. For radical-radical reactions, additionally, we used the broken-(spin)-symmetry \textit{ansatz} to correctly describe the open-shell singlet state of the system \citep{neese_2004,abe2013}, allowing two unpaired electrons with opposite spin to occupy different orbitals, to avoid forcing the recombination between the two radicals. The stability of the wave function so obtained was verified prior to the geometry optimisation step. This methodological strategy was also verified by some of us for the treatment of the electronic structure of biradical systems \citep{Enrique-Romero2022}, including a comparison of the broken symmetry approach with CASPT2 calculations \citep{Enrique-romero2020}.

The computational requirements needed to perform single energy point calculations at the DLPNO-CCSD(T)-F12 level of theory on open-shell systems exceeded our resources, so they were calculated at the B3LYP-D3(BJ)/ma-def2-TZVP level of theory. However, DLPNO-CCSD(T)-F12 calculations were possible for closed-shell systems (reactions \ref{eq:neu-neu} and \ref{eq:salt}), and demonstrated the reliability of the DFT methodology. 

Additionally, we calculated the unimolecular rate constant and half-life time for reaction \ref{eq:rad-rad} using the classical transition state theory, following the Eyring's equation:
    \begin{equation}
        k(T) = \kappa \frac{k_B T}{h} e^\frac{-\Delta G \ddagger}{RT}
    \end{equation}
where $\kappa$ is the transmission coefficient (which in our case is assumed to be 1 since tunnelling effects are not operative due to the non-participation of light atoms like H but heavy entities), $k_B$ is the Boltzmann constant, \textit{T} is the temperature, \textit{h} is the Planck constant, $\Delta$G$\ddagger$ is the Gibbs free energy barrier (computed at a given temperature) and \textit{R} is the gas constant. Once obtained the rate constant, one can compute the half-life time of the reactants through the equation $ t_{1/2} (T) = ln(2)/k(T)$, which is defined as the time for the reactants to be half consumed.

\begin{figure}[htb]
\centering
	\includegraphics[width=0.5\linewidth]{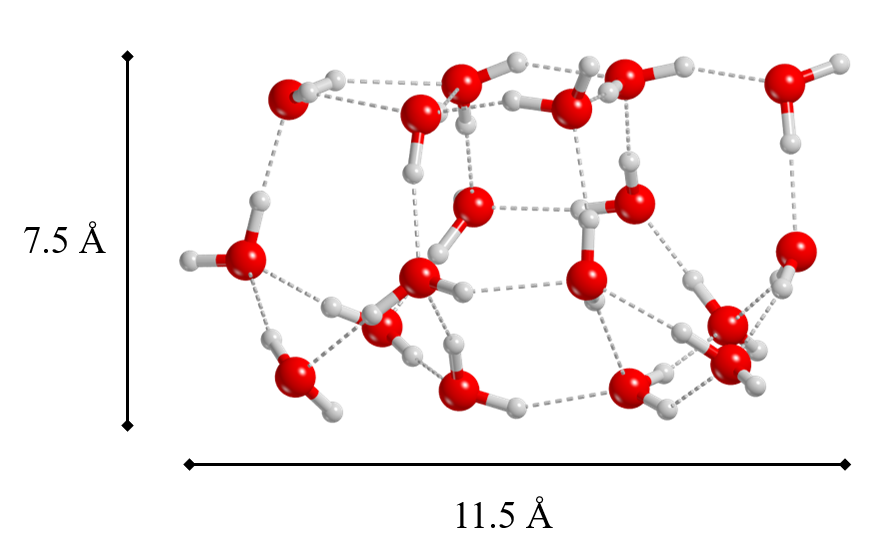}
\caption{The 18 water molecules cluster, optimized at the B3LYP-D3(BJ)/ma-def2-TZVP level of theory.}
\label{fig:icemodel}
\end{figure}

Finally, our water ice cluster model consists of 18 H$_2$O molecules (see Figure \ref{fig:icemodel}) and has been used in previous studies by some of us \citep{rimola2014,rimola2018,perrero2022,enrique-romero2019}. The structure represents a compact, amorphous, and flat water ice surface, whose size is a compromise between a more realistic extended model and one that allows to adopt a high level of theory at a reasonable computational cost.

\section{Results and discussion} 

\begin{figure}[htb]
\centering
	\includegraphics[width=0.7\linewidth]{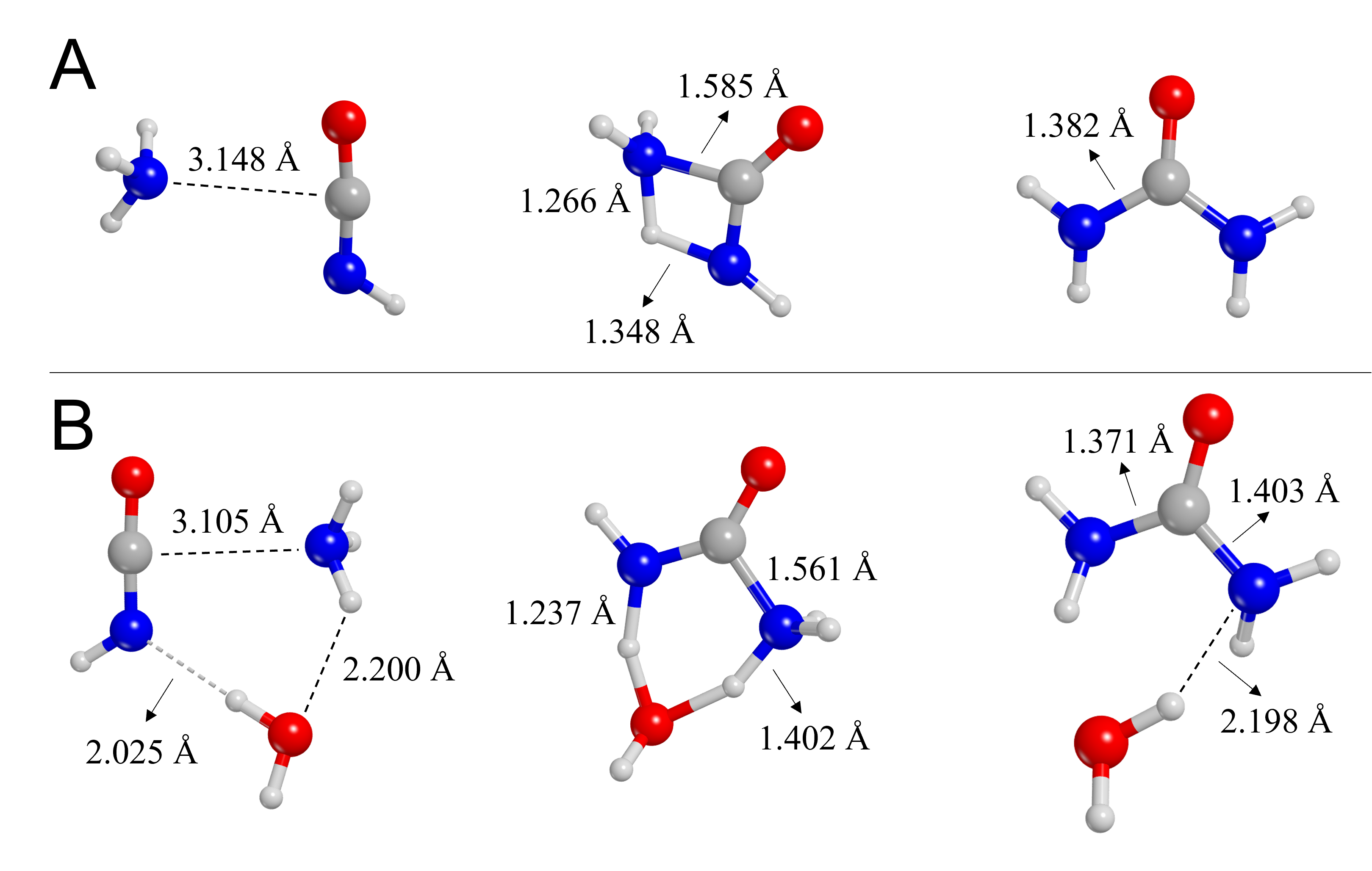}
\caption{Structures (from left to right) of reactant, transition state and product for the gas phase reaction in absence (A) and in presence (B) of one water molecule. Structures are optimized at B3LYP-D3(BJ)/ma-def2-TZVP level of theory, bond distances are in \AA.}
\label{fig:benchmark}
\end{figure}

As mentioned above, we first characterised the reaction between HNCO and \ch{NH3}, in the gas phase. Because these molecules are closed-shell, we expect to find a high energy barrier for the formation of urea, which is confirmed by our calculations. The gas-phase reaction between isocyanic acid and ammonia takes place in a concerted fashion and goes through the formation of the \ch{C-N} bond simultaneously to a proton transfer from the nitrogen atom of \ch{NH3} to that of HNCO (see Figure \ref{fig:benchmark}A). The transition state structure presents a four-membered ring, which is a highly strained geometry, and hence the energy barrier is as high as 136 kJ mol$^{-1}$ (at B3LYP-D3(BJ) level of theory). By introducing one water molecule into the reaction, the transition state structure presents a six-membered ring. This is a geometrically less strained structure because of the participation of the added water molecule in the proton transfer. That is, the water molecule assists the H-transfer by receiving a proton from ammonia and donating one of its protons to the NH moiety of isocyanic acid (see Figure \ref{fig:benchmark}B). This water-assisted proton transfer mechanism has a large impact on the energy barrier, as it decreases to 46 kJ mol$^{-1}$ (at B3LYP-D3(BJ) theory level). 

\begin{table*}[cols=7,pos=h]
\caption{Gas phase benchmark. Optimised geometries and the ZPE correction are computed at the B3LYP-D3(BJ)/ma-def2-TZVP level of theory. Energies in kJ mol$^{-1}$.}\label{tab:benchmark}
\begin{tabular*}{\tblwidth}{@{} LCCCCCC@{} }
\toprule
& \multicolumn{3}{c}{Electronic energies} & \multicolumn{3}{c}{Enthalpies} \\
HNCO + \ch{NH3} & B3LYP-D3(BJ) &	DLPNO-F12 &	CCSD(T)-F12 & B3LYP-D3(BJ) &	DLPNO-F12 &	CCSD(T)-F12 \\
\midrule
Reactant & 0.0 &    0.0   &   0.0 & 	0.0 &	0.0 &	0.0 \\
Transition state & 136.0 & 	136.5 & 	135.2 & 	140.1 &	140.6 &	139.2 \\
Product & -75.9 & 	-80.4	 & -81.1 & 	-57.9 &	-62.4 &	-63.1 \\
\midrule
HNCO + \ch{NH3} + \ch{H2O} & B3LYP-D3(BJ) &	DLPNO-F12 &	CCSD(T)-F12 & B3LYP-D3(BJ) &	DLPNO-F12 &	CCSD(T)-F12 \\
\midrule
Reactant & 0.0 &    0.0   &   0.0 & 0.0 &	0.0 &	0.0 \\
Transition state & 46.3 &	53.3 &	51.9 & 49.9 &	56.9 &	55.5 \\
Product & -75.4 &	-81.3 &	-81.8 & -58.6 &	-64.5 &	-65.0 \\
\bottomrule
\end{tabular*}
\end{table*}

As mentioned in Section \ref{sec:method}, these closed-shell/closed-shell processes have also been used to perform a benchmark study to calibrate the adequacy of the B3LYP-D3(BJ)/ma-def2-TZVP level of theory. The results are summarised in Table \ref{tab:benchmark} and, as can be seen, indicate
a very good performance of B3LYP-D3(BJ)/ma-def2-TZVP compared to CCSD(T)-F12. The relative errors of the potential energy barriers at B3LYP-D3(BJ) with respect to CCSD(T)-F12 are 0.6\% and 11\% in the absence and in the presence of the water assisting molecule, respectively. The energies computed at DLPNO-CCSD(T)-F12 slightly improve the quality reached by B3LYP-D3(BJ), with relative errors of
1\% and 3\% in absence and in the presence of water with respect to CCSD(T)-F12.

Given the previous results, we simulated the same reaction on the water ice cluster model, since a larger number of water molecules assisting the proton transfer step could reduce the energy barrier even more.  

\begin{figure*}[htb]
\centering
	\includegraphics[width=\linewidth]{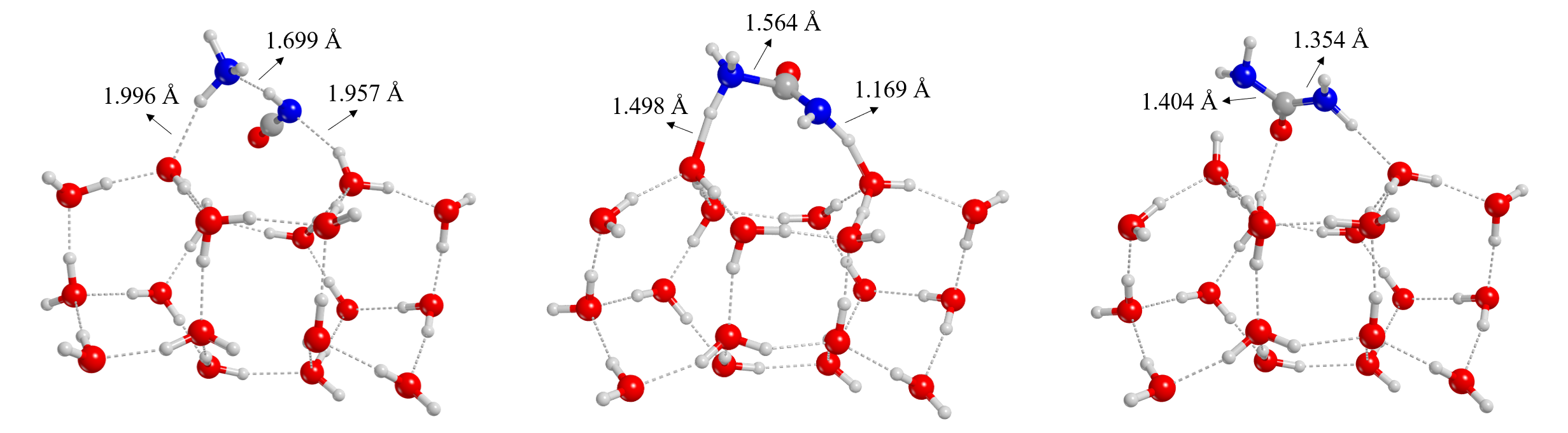}
\caption{B3LYP-D3(BJ) optimised structures for the reactant (left), transition state (centre), and product (right) of the \ch{HNCO + NH3} -> \ch{NH2CONH2} reaction on the water ice cluster model.}
\label{fig:neutral-neutral}
\end{figure*}

The adsorption geometry of HNCO and \ch{NH3} on the flat surface of the cluster does not seem to point towards the formation of urea, since we found a number of complexes characterised by a H-bond between the hydrogen of HNCO and the nitrogen of \ch{NH3}. Most likely, this orientation yields ammonium cyanate, rather than urea. Additionally, we also found some reaction pathways in which the formation of urea occurred without involving the water molecules at the surface, with transition states resembling four-membered ring structures. 
However, we found a mechanism involving icy water molecules in the proton transfer event similarly to the reaction in the presence of one water molecule (see Figure \ref{fig:neutral-neutral}).
Here, to form the \ch{C-N} bond, a subtle rotation of the HNCO is needed to bring the C atom close to ammonia, and four water molecules participate in the proton transfer, which takes place through the network of H-bonds established in the reactant complex. However, the strong interaction between the reactants and the ice is against the reaction, which presents a barrier of 97.8 kJ mol$^{-1}$, significantly higher than that obtained in the presence of one water molecule (46 kJ mol$^{-1}$). Therefore, the large number of water molecules offered by the surface improves the proton transfer mechanism with respect to the gas-phase process, but the strong interaction of the adsorbates with the ice seems to be an even greater drawback.

\begin{figure*}[htb]
\centering
	\includegraphics[width=\linewidth]{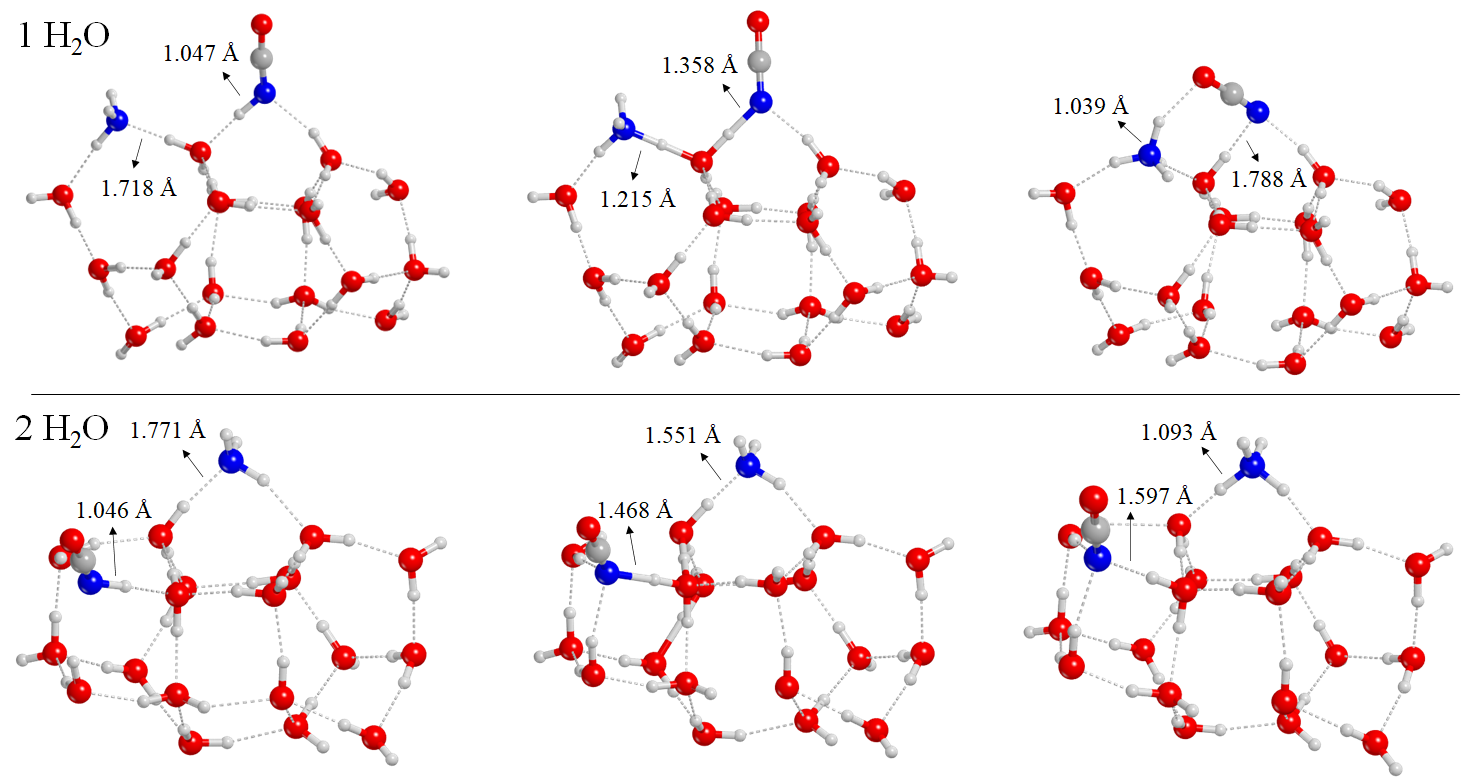}
\caption{B3LYP-D3(BJ) optimised structures for the reactant (left), transition state (centre), and product (right) of the  \ch{HNCO + NH3 -> NH4+ OCN-} reaction mediated by one (top) and two (bottom) water molecules.}
\label{fig:salt}
\end{figure*}

None of the \ch{HNCO + NH3 + ice} adsorption complexes yielded the spontaneous formation of the \ch{NH4+ OCN-} ion pair. Furthermore, geometry optimisations considering the ion pair already formed as initial structures evolved towards the formation of HNCO and NH$_3$, meaning that complexes in which the NH$_4$$^+$ and OCN$^-$ ions interact directly are not stable. On the other hand, the water molecules of the ice surface can stabilise the charges of the two isolated ions, due to H-bonds. Therefore, we searched for adsorption geometries in which HNCO and NH$_3$ were separated by some water molecules to avoid their direct interaction and, at the same time, interact with them, thus favouring the formation and survival of the \ch{NH4+ OCN-} ion pair.
We found two structures in which HNCO and \ch{NH3} are separated, by one and two water molecules, respectively (see Figure \ref{fig:salt}). Thus, we computed the formation of the ion pair in which the proton transfer takes place through the H$_2$O molecules that separate HNCO and \ch{NH3}. In both cases, the processes are not spontaneous and present a barrier of 19.0 and 28.8 kJ mol$^{-1}$ when one and two \ch{H2O} molecules are involved, respectively. Moreover, the reactions are endothermic and hence the processes are thermodynamically disfavored.  

In relation to the formation of the \ch{NH4+ OCN-} ion pair, several studies are available in the literature. 
While both \cite{raunier2003nh4ocn} and \cite{theule2011} found a barrier for the dissociation process of HNCO in water ice, \cite{raunier2004nh4onc} found that the formation of \ch{NH4+ OCN-} is spontaneous when HNCO is embedded in a mixture of NH$_3$:H$_2$O ice, meaning that the formation of the \ch{NH4+ OCN-} ion pair is more favourable than the formation of \ch{OCN-} in pure water. On the other hand, \cite{mispelaer2012} studied the formation of \ch{NH4+ OCN-} by thermally treating a mixture of HNCO and NH$_3$ ice and found a small barrier for the process (0.4 kJ mol$^{-1}$, in Table \ref{tab:reactionsw18}). From the calculations of \cite{raunier2004nh4onc} it appears that the barrierless ion pair formation is due to the solvation of both HNCO and NH$_3$ on behalf of three or more water molecules. Our results are in line with all these findings, because from previous work it seems clear that the spontaneous formation of the \ch{NH4+ OCN-} ion pair can take place in the bulk of the ice, where the number of water molecules interacting with the reactants is larger than those offered by the surface. Indeed, on the latter, the species tend to be adsorbed and do not experience bulk water solvation effects, and accordingly the stabilisation inferred by the surface molecules is not sufficient to guarantee the spontaneous formation of the ion pair. 

\begin{table}[cols=6,pos=ht]
\caption{Enthalpy variations at 0 K (in kJ mol$^{-1}$) of the reactions modeled on the water ice cluster at B3LYP-D3(BJ)/ma-def2-TZVP level of theory, compared with literature data. References: a) \cite{mispelaer2012}, b) \cite{raunier2004nh4onc}, c) \cite{slate2020}. }\label{tab:reactionsw18}
\begin{tabular*}{\tblwidth}{@{} LCCLCC@{} }
\toprule
Literature & Barrier & Reference & Present work & $\Delta$H$_{TS}$ & $\Delta$H$_{prod}$ \\
\midrule
  &  &  & \ch{HNCO + NH3 -> NH2CONH2} & 97.8  & -28.8   \\
  \ch[arrow-offset=.0em]{HNCO + NH3 ->[$\Delta$] NH4+ OCN-} & 0.4 &   experiments (a) & \ch[arrow-offset=.0em]{HNCO + NH3 ->[1 H2O] NH4+ OCN-} & 19.0 & 8.0     \\
  \ch{HNCO + NH3 -> NH4+ OCN- } ($\dagger$) & no &   theory (b) & \ch[arrow-offset=.0em]{HNCO + NH3 ->[2 H2O] NH4+ OCN-} & 28.8 & 34.5   \\
& & & \ch{HNCO + NH2 -> NHCONH2}  & 78.8  & -21.7    \\
 \ch{H2NCO + NH2 -> NH2CONH2} & 4.0 & theory (c) & \ch{H2NCO + NH2 -> NH2CONH2} & 3.9 & -357.2   \\
& & & \ch{H2NCO + NH2 -> NH3 + HNCO} &  $\approx$ 20.0  &  -290.9  \\
\bottomrule
\end{tabular*}
\begin{tabular}{L}
    $^{\dagger}$ The reaction is performed on a 3 H$_2$O cluster.
\end{tabular}
\end{table}

Interestingly, the reactions studied so far involve closed-shell species, and hence we were able to compute the potential energy barriers at DLPNO-CCSD(T)-F12 level of theory. By comparing these coupled cluster results with those obtained at B3LYP-D3(BJ) level (see Table \ref{tab:b3lyp-dlpno}), we can see that B3LYP-D3(BJ) slightly underestimates the barriers with an average error of about 16\%. In view of this fairly agreement between the two methods, we proceeded with the rest of the calculations at the B3LYP-D3(BJ) theory level. The remaining reactions involve open-shell species that increase their computational cost. 

\begin{table}[cols=3,pos=h]
\centering
\caption{Potential energy barriers ($\Delta$E$_{TS}$, in kJ mol$^{-1}$) of the closed-shell reactions performed on the water ice cluster model. Energies obtained with B3LYP-D3(BJ)/ma-def2-TZVP (DFT) are compared with single point energy calculations at DLPNO-CCSD(T)-F12 (DLPNO) level of theory.}
\label{tab:b3lyp-dlpno}
\begin{tabular*}{0.5\columnwidth}{@{} LCC @{}}
\toprule
Reaction & DFT & DLPNO \\
\midrule
 \ch{HNCO + NH3 -> NH2CONH2} & 106.4   & 116.6     \\
 \ch[arrow-offset=.0em]{HNCO + NH3 ->[1 H2O] NH4+ OCN-}  & 31.6  & 39.2   \\
 \ch[arrow-offset=.0em]{HNCO + NH3 ->[2 H2O] NH4+ OCN-}  & 40.2  & 50.8   \\
\bottomrule
\end{tabular*}
\end{table}

We characterised the two proposed alternative pathways involving radical species, in a way similar to \cite{slate2020}, that is, reactions \ref{eq:rad-neu} and \ref{eq:rad-rad}. Considering the enhanced reactivity of radicals, we expect the barriers of these processes to be lower than those involving only the closed-shell species. 

\begin{figure*}[htb]
\centering
	\includegraphics[width=0.8\linewidth]{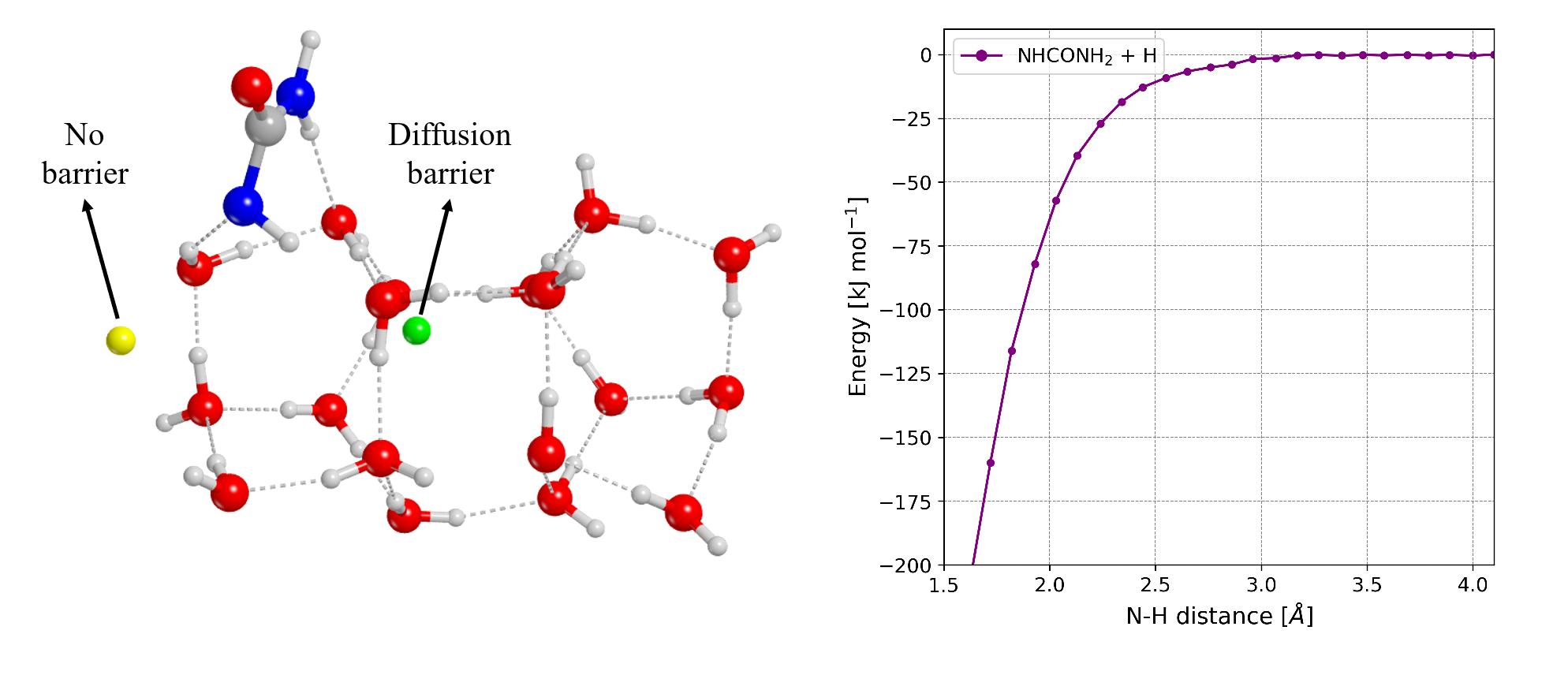}
\caption{Left panel: there are two possible initial positions for H atoms (represented as yellow and green atoms) through which the H-addition to NH$_2$CONH can take place. The yellow position evolves spontaneously to form urea, whereas the green position does not. Right panel: distinguished potential energy profile obtained by scanning the N-H distance from the green-represented H atom.}
\label{fig:hydrogenation}
\end{figure*}

Firstly, we characterised reaction \ref{eq:rad-neu}, which is between HNCO and the \ch{NH2} amino radical, yielding \ch{NHCONH2}. The calculated energy barrier is 78.8 kJ mol$^{-1}$, only 20 kJ mol$^{-1}$ less than the closed-shell/closed-shell reaction. The reactants are tightly bound (well-stabilised) to the surface, and hence the energy barrier is prohibitive in the ISM. We also noticed that, in this reaction, no proton transfer takes place, therefore the water ice is uniquely participating as reactant concentrator/supplier and, in view of the released reaction energy (-21.7 kJ mol$^{-1}$), as energy dissipator. In order to finally form urea, a final H-addition to \ch{NHCONH2} is necessary. This process can be barrierless depending on the orientation of the two partners. If the H atom is not pointing towards the NH moiety, it has to overcome a small diffusion barrier to react, which moreover can occur via quantum tunnelling. In agreement with that, in this final H-addition, by optimising the initial states in the triplet electronic state, we found two different positions for the H atoms (see the yellow- and green-represented H atom in Figure \ref{fig:hydrogenation}, left panel). When changing to the singlet electronic state, the yellow H atom yielded the product barrierlessly, but this is not the case for the green one. For this case, to assess the energetics of the H-addition, we performed a scan calculation along the \ch{N-H} distance (Figure \ref{fig:hydrogenation}, right panel), in which the potential energy is almost constant at a \ch{N-H} distance between 3 \AA~and 4 \AA~(i.e., when the hydrogen atom is approaching the reactive site), and falls down into the potential well at shorter distances, this way forming the product. This indicates that, although the initial state is stable and apparently does not react spontaneously, in practice the hydrogenation reaction evolves toward the formation of urea in a non-energetic way.

\begin{figure}[htb]
\centering
	\includegraphics[width=0.7\linewidth]{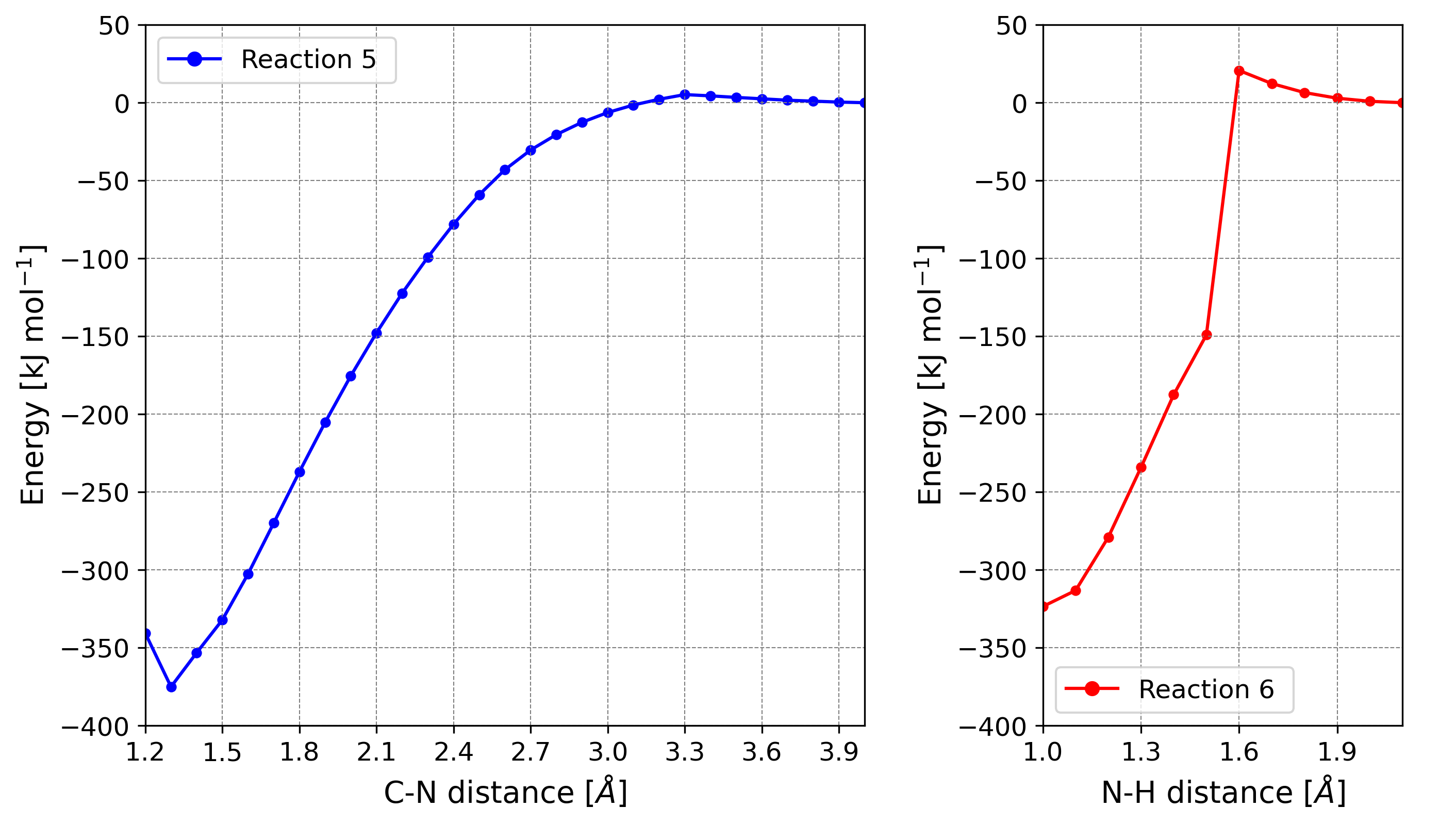}
\caption{Scan profile for reactions \ref{eq:rad-rad} and \ref{eq:rad-rad-reagent} in which the energy along the reaction coordinate (the C-N distance in the first case and the N-H distance in the second case) is followed. The reaction that yields HNCO + NH$_3$ (reaction \ref{eq:rad-rad-reagent}) has a higher barrier to overcome to yield the product with respect to the formation of urea (reaction \ref{eq:rad-rad}).}
\label{fig:comparison}
\end{figure}

Finally, the radical-radical reaction between \ch{H2NCO} and \ch{NH2} was investigated. It presents a much lower energy barrier compared to the other two reactions, but it has a competitive channel that yields HNCO + \ch{NH3} through an H-abstraction. The occurrence of one process or the other depends on the geometry of the adsorption complex. This was also found in \cite{Enrique-Romero2022} by studying a well-suited pair of radical-radical couplings forming different interstellar complex organic molecules. From a thermodynamic point of view, reaction \ref{eq:rad-rad} is more exothermic than reaction \ref{eq:rad-rad-reagent}. In relation to the energy barriers, we performed scan calculations along the \ch{C-N} (forming urea) and the \ch{H-N} (forming HNCO + \ch{NH3}) distances to estimate and compare the potential energy profiles of the two reactions. In Figure \ref{fig:comparison} we observe that the highest energy point of reaction \ref{eq:rad-rad} reaches $\approx$ 5 kJ mol$^{-1}$, while for reaction \ref{eq:rad-rad-reagent} it is located at $\approx$ 20 kJ mol$^{-1}$. 
For the former reaction, we were able to localise the transition state structure, presenting an actual ZPE-corrected energy barrier of 3.9 kJ mol$^{-1}$. For the latter reaction, any attempt to localise the actual transition state structure failed, probably because of hysteresis effects caused by the high mobility of the H atom. Despite this, the distinguished energy profiles clearly indicate that the radical-radical coupling is, in terms of energy barriers, more favourable than the H-abstraction. Thus, as shown in these results (alongside the thermodynamics), the formation of urea seems to be the most favourable path through this radical-radical coupling path. However, the H-abstraction can also be facilitated by tunnelling effects so that this competitive channel cannot be totally excluded and should not be overlooked.

In reaction \ref{eq:rad-rad}, the ice surface does not actively participate in the mechanism, but (as postulated in the closed-shell/radical reaction) it encompasses the role of reactant concentrator/supplier and third body.

\begin{figure}[htb]
\centering
	\includegraphics[width=0.45\columnwidth]{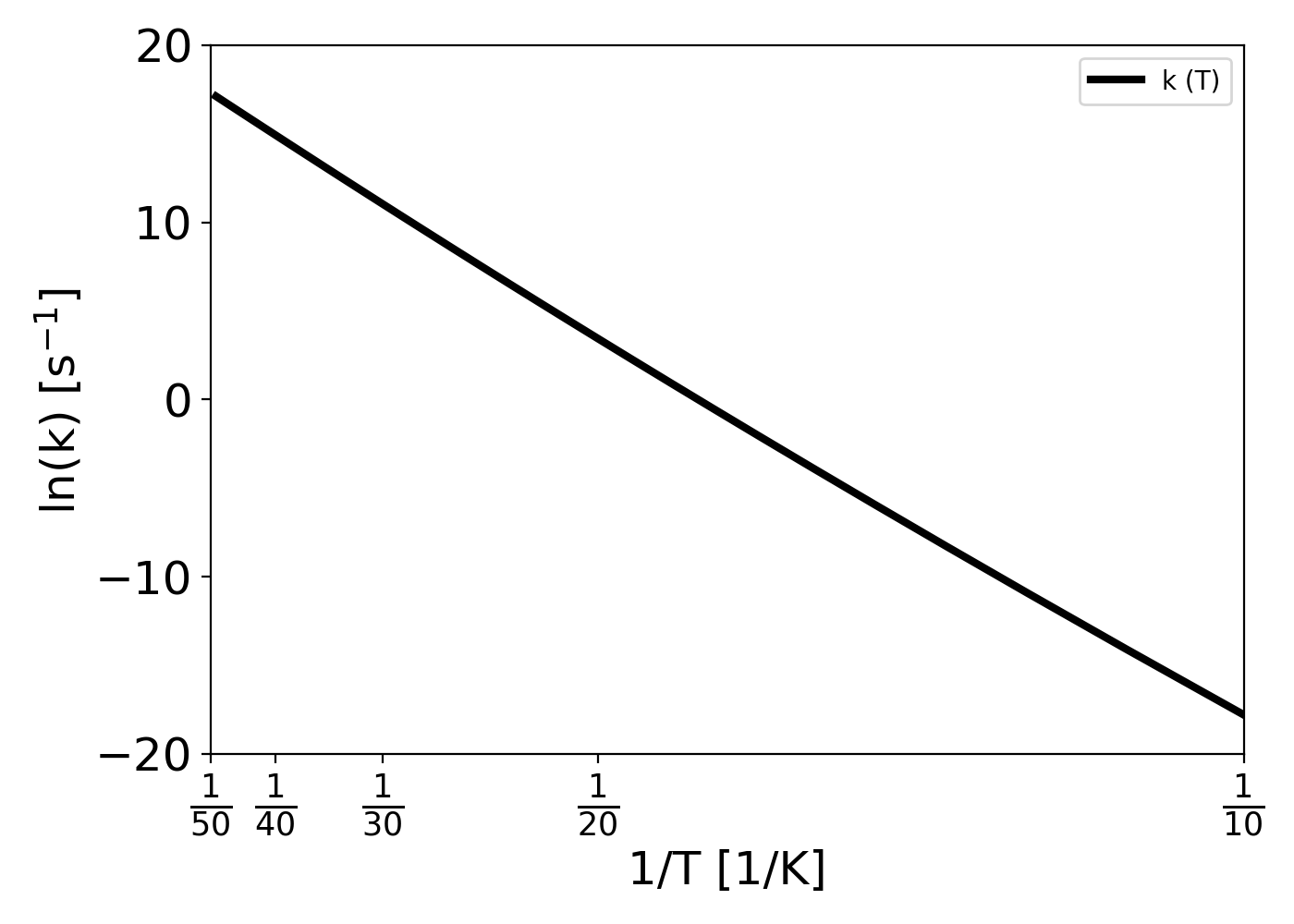}
\caption{Arrhenius plot of the rate constant (in s$^{-1}$) obtained with the classical Eyring equation for reaction \ref{eq:rad-rad} between 10 and 50 K. }
\label{fig:kinetics}
\end{figure}

\begin{table}[cols=3,pos=ht]
\caption{Unimolecular rate constants (in s$^{-1}$) and half-life times (in s) computed for reaction \ref{eq:rad-rad} between 10 and 50 K with the classical Eyring equation. }
\label{tab:half-life}
\begin{tabular*}{0.4\columnwidth}{@{} LLL @{}}
\toprule
T (K) & 	k (s$^{-1}$)	& t$_{1/2}$  (s) \\
\midrule
10 &	8.0 $\times$ 10$^{-10}$ &   8.6 $\times$ 10$^{8}$ \\
15 &	7.5 $\times$ 10$^{-3}$ &    9.2 $\times$ 10$^{1}$ \\
20 &	2.5 $\times$ 10$^{1}$ &     2.8 $\times$ 10$^{-2}$ \\
25 &	3.4 $\times$ 10$^{3}$ & 	2.0 $\times$ 10$^{-4}$ \\
30 &	9.5 $\times$ 10$^{4}$ & 	7.3 $\times$ 10$^{-6}$ \\
35 &	1.0 $\times$ 10$^{6}$ & 	6.6 $\times$ 10$^{-7}$ \\
40 &	6.5 $\times$ 10$^{6}$ & 	1.1 $\times$ 10$^{-7}$ \\
45 &	2.7 $\times$ 10$^{7}$ & 	2.5 $\times$ 10$^{-8}$ \\
50 &	8.7 $\times$ 10$^{7}$ & 	8.0 $\times$ 10$^{-9}$ \\
\bottomrule
\end{tabular*}
\end{table}

The radical \ch{H2NCO} can be obtained from the reaction of CN + \ch{H2O} \citep{rimola2018}, but is also the radical of formamide, as suggested by \cite{slate2020}. In the latter work, the authors found a barrier of 4 kJ mol$^{-1}$ for reaction \ref{eq:rad-rad}, which is in agreement with our results (see Table \ref{tab:reactionsw18}). 
However, the authors also argued that the positive outcome of the reaction depends on the chance of an electronic spin change of the system because the most stable electronic state is the (unreactive) triplet, at variance with the less stable (reactive) singlet. This also holds for our simulations. However, during the long-life time of interstellar molecular clouds, events like photon and/or cosmic ray incidence and thermal/shock waves can indeed induce triplet-to-singlet electronic state changes, this way enabling the occurrence of the radical-radical couplings.

Assuming this, we performed a kinetic study on reaction \ref{eq:rad-rad} to know at which temperature the reaction can overcome the ZPE-corrected energy barrier of 3.9 kJ mol$^{-1}$. We computed the $\Delta G \ddagger$ from 10 to 50 K and we obtained the unimolecular rate constant of the reaction (see Figure \ref{fig:kinetics}), together with the half-life time of the reactants (see Table \ref{tab:half-life}). At a temperature of 10 K, it takes 8.6 $\times$ 10$^8$ s (corresponding to 27 years) for the reactant to be half consumed, so the reaction is very slow. However, at 20 K the half-life time becomes 0.03 s, indicative of a very fast reaction, which enhances its velocity as the temperature increases. Therefore, we can conclude that a barrier of 3.9 kJ mol$^{-1}$ does not hamper this reaction under ISM conditions. However, a factor that should be considered is the availability and proximity of the radicals in order to yield the product.

\section{Conclusion}		
In this work, we investigated the formation of urea on an 18 H$_2$O molecules ice cluster through different chemical reactions that involve both closed-shell and open-shell species. We first characterised the reaction between HNCO and NH$_3$ (closed-shell/closhed-shell) in absence and in presence of one H$_2$O molecule, with the aim of testing the accuracy of the B3LYP-D3(BJ)/ma-def2-TZVP methodology by comparing it with the reference CCSD(T)-F12. We then characterised the same reaction on the 18 \ch{H2O} molecules cluster model, finding that although water ice participates in the proton transfer step necessary to yield urea (hence exerting some catalytic effects), the energy barrier of the process is $\approx$ 100 kJ mol$^{-1}$, hampering the reaction to take place in the ISM. We also studied a radical/closed-shell (HNCO + NH$_2$) and a radical-radical (H$_2$NCO + NH$_2$) pathways, to assess if they were a more favourable process. 
We summarise our findings for these open-shell reactions in the following: (i) the radical/closed-shell reaction has an energy barrier of about $\approx$ 80 kJ mol$^{-1}$, and therefore it can also be discarded in the ISM; (ii) the radical-radical reaction is almost barrierless, but it has a competitive channel (H-abstraction forming HNCO + NH$_3$) which, however, is characterised by an energy barrier of $\approx$ 20 kJ mol$^{-1}$ (although tunnelling effects can be operative); (iii) in view of the nature of the reactions and their large and negative reaction energies, the role of the icy grains in these reactions is of reactant concentrators/suppliers and energy dissipators.
In the closed-shell/closed-shell reaction, the water ice surface participates in the proton transfer, while in the radical/closed-shell and radical-radical reactions it simply binds the reactants. In all reactions, the strong interaction of HNCO and \ch{NH3} with the ice surface can represent an obstacle to the course of the reaction.
As some authors suggest \citep{brigiano2018,slate2020,kerkeni2023}, in addition to radical-radical reactivity, charged pathways also represent a promising mechanism to explain the formation of complex molecules such as urea. Their activation barriers are generally lower than those of reactions between closed-shell species, making these pathways feasible in the cold and rarefied environments of the ISM. In conclusion, the presence of the ice would be crucial also in this case, especially to provide a third body able to dissipate the energy liberated by the reaction of energetic species like cations.

\section*{Declaration of Competing Interest}
The authors declare that they have no known competing financial interests or personal relationships that could have appeared to influence the work reported in this paper.

\section*{Acknowledgments}
This project has received funding within the European Union’s Horizon 2020 research and innovation programme from the European Research Council (ERC) for the project ``Quantum Chemistry on Interstellar Grains” (QUANTUMGRAIN), grant agreement No 865657. The authors acknowledge funding from the European Union’s Horizon 2020 research and innovation program Marie Sklodowska-Curie for the project ``Astro-Chemical Origins” (ACO), grant agreement No 811312. MICINN (project PID2021-126427NB-I00) is also acknowledged. CSUC supercomputing center is acknowledged for allowance of computer resources.

\section*{Data Availability}
The data underlying this article are freely available in Zenodo at \url{https://doi.org/10.5281/zenodo.8285735}.



\bibliographystyle{model2-names}

\end{document}